\documentstyle[12pt]{article}
\font\titlefont=cmbx10 scaled \magstep3
\begin{document}
\begin{center}
{\titlefont The Limits of Information }
 
\vskip .3in
Jacob D. Bekenstein\footnote{email: bekenste@vms.huji.ac.il} \\
\vskip .1in
Racah Institute of Physics,
Hebrew University of Jerusalem\\
Jerusalem 91904 ISRAEL\\
\end{center}
\abstract{Black holes have their own thermodynamics including notions of
entropy and temperature and versions of the three laws.  After a light
introduction to black hole physics, I recollect how black hole thermodynamics
evolved in the 1970's, while at the same time stressing conceptual points
which were given little thought at that time, such as why the entropy should
be linear in the black hole's surface area.  I also review a variety of
attempts made over the years to provide a statistical mechanics for black hole
thermodynamics.  Finally, I discuss the origin of the information bounds for
ordinary systems that have arisen as applications of black hole
thermodynamics.
\smallskip

\noindent
Keywords: Black holes, entropy, second law, information, information bound.
}
\vskip .1in
\leftskip 4cm
{\it A theory is the more impressive the greater the  simplicity of its
premises, the more different kinds of things it relates, and the more
extended its area of applicability.  Therefore, the deep impression which
classical thermodynamics made upon me.  It is the only physical theory of a
universal content concerning which I am convinced that within the framework
of applicability of its basic concepts, it will never be overthrown . . . .}

\smallskip
\noindent
A. Einstein,
{\it Autobiographical Notes}
%\newpage

\leftskip=0.0cm
\baselineskip=14pt

\bigskip
\centerline{\bf Introduction}
\medskip

Although the citation from Einstein captures the professed attitude of many
physicists, many others would today regard it as a basically sentimental
statement.  For we have become accustomed to regard thermodynamics as a
straight consequence of statistical mechanics and the atomic hypothesis.  The
usual paradigm, inferred from myriad examples - ideal gas, black body
radiation, a superfluid, etc. - is that a system made up of a
multitude of similar parts - molecules, electrons, phonons - with weak
interactions and with no initial correlations between constituents
(Boltzmann's {\it Stosszahl-Ansatz}), will automatically exhibit thermodynamic
behavior.   Thus - so the claim - it is statistical mechanics, not
thermodynamics, which is the theory of ``universal content''.  The advent of
black hole thermodynamics thirty years ago seems, however, to have turned the
tables on this ``modern'' assessment of thermodynamics' secondary status.

In effect, black holes provide a second paradigm of thermodynamics.  Black
hole thermodynamics has meaning already at the classical level.  It possesses a
first law (conservation of energy, of momentum, of angular momentum and of
electric charge), as well as a second law (in a generalized version) and a
third law which delimits the kingdom of black holes.  Black hole
thermodynamics is no ordinary thermodynamics.  Gravitation is all important
in it, while in traditional thermodynamics gravitation is a nuisance which is
customarily ignored.  Information about a black hole's interior is not just
practically unavailable, as in the garden variety thermodynamic system;
rather, there are physical barriers forbidding acquisition of
information sequestered in a black hole.  And, again unlike everyday
thermodynamic systems, a black hole is a monolithic systems with no parts. 
Granted, as the usual description has it, a black hole is usually the product
of the collapse of ordinary, thermodynamic, matter, but that infallen matter
becomes invisible and thermodynamically irrelevant.  To top it off, we are not
sure today, despite vociferous claims to the contrary, whether there exists a
statistical mechanics which reduces to black hole thermodynamics in some
limit.  Einstein would have been pleased: thermodynamics seems to stand by
itself in the black hole domain.

And the issue is not just academic.  Black hole thermodynamics seems to tell
us things about mundane physical systems.  For instance, a very pragmatic
question in modern technology is how much information can be stored by {\it
whatever\/} means in a cube of {\it whatever\/} composition one centimeter on
the side.  As we shall see, black hole thermodynamics has suggested upper
bounds on this quantity; these are the limits of information.

\bigskip\centerline{\bf  What is a Black Hole ?}
\medskip

Ask what is a black hole and you will get many answers. The purist will say:
a solution of Einstein's theory of gravity - general relativity - representing
a spacetime which over most of its extension is like the one we are familiar
with from special relativity, but which includes a region of finite spatial
extent whose interior and boundary are totally invisible from the rest of the
spacetime.  Colloquially that invisible region is a black hole.  Others will
define the black hole differently: a tear in the fabric of spacetime or a
gravitational soliton.  The first of these emphasizes the familiar notion that
whatever goes into a black hole cannot return. The second hints at the
property that that invisible region behaves in many ways like an ordinary
object: it is localized, can move, can be scattered, attracts other objects,
etc.

It is fairly certain that a black hole can form from the collapse of an old
star; the X-ray sources observed in our galaxy since the 1970's include a
class of some dozens of rapidly flickering ones, each of which is thought to
harbor a black hole of stellar mass.  It is also fairly certain that quasars,
those extremely luminous beacons in distant parts of the universe, of which
some 5000 have been catalogued, represent active regions in galaxies, each of
which is energized by a very massive black hole at its center.  Our own Milky
Way, although no quasar, is known to have a modestly massive black hole at its
core.  Finally, it is suspected that microscopic scale black holes formed in
the universe when it was extremely dense from the strong density fluctuations
which would naturally arise then.  None of these primordial black holes have
yet been spotted.  However, what the above list emphasizes is that black holes
may be found with a wide range of masses, from $10^{15}$ g for the primordial
ones to $10^{41}$ g for those in quasars, and perhaps higher.

Much of what we know about the physics of black holes is based on exact
solutions to Einstein's theory, the Schwarzschild, Reissner-Nordstr\"om,
Kerr, and Kerr-Newman solutions (Misner, Thorne and Wheeler 1973).  The most
general of these, the Kerr-Newman solution, represents a black hole possessing
mass $m$, electric charge $q$ and angular momentum $j$.  The black hole can
move, and so it can also have linear momentum, but I shall gloss over this
possibility here.  One feature of the black hole phenomenon which is clear
from the Kerr-Newman solution is the (event) horizon, the boundary of the
invisible region.  The warping of spacetime, archetypical of Einstein's
description of gravity, causes the local lightcones familiar from special
relativity to tilt so that their exterior boundary lies tangent to the
horizon.  Thus, in Wheeler's words (Ruffini and Wheeler 1971), the horizon
acts like a ``one-way membrane'': no object or light ray that crosses it
inward-going can ever recross it outwardly.  For in special relativity language
that would be tantamount to moving faster than light.  As a result of this
purely geometric obstacle, information that enters the black hole is
permanently trapped inside the horizon, and is unrecoverable to any observer
outside the black hole precisely because no signals, whatever their nature,
can cross the horizon outward.  A horizon constitutes a barrier to information
flow.

How big is the horizon ?  From Schwarzschild's solution representing a
spherical black hole with exactly one parameter, mass $m$, we learn that its
radius is $R_h=2Gmc^{-2}$ where $G$ is Newton's constant and $c$ the speed of
light.  For a solar mass black hole $R_h\approx 3$ km.  From the three
parameters, $m$, $q$ and $j$, of the more general (an aspherical) Kerr-Newman
black hole (KNBH)  we can form three lengths: $M=Gm c^{-2}$, $Q=\surd Gqc^{-2}$
and $a=jm^{-1} c^{-1}$.  In terms of these the area of the horizon $A$ is given
by the important formula
\begin{equation}
A=4\pi(R_+^2+a^2);\quad R_+\equiv M+\sqrt{M^2-Q^2-a^2}
\label{area}
\end{equation}
From this one can get an idea of the generic size of the horizon; for $m$ a
solar mass, $(A/4\pi)^{1/2}$ is always of order 1 km. 

When charge $\delta q$ is added to a KNBH, its parameter $q$
grows by $\delta q$; likewise when angular momentum $\delta j$ is added, $j$
grows by $\delta q$.  And $m$ grows by whatever energy (and work) was added. 
The laws of energy, charge and angular momentum retain their usual meaning. 
Not so other respected laws of the physicist.  For instance, baryon number
conservation is `transcended' by black holes (Ruffini and Wheeler 1971;
Bekenstein 1972a).  Addition of baryon number to a KNBH is
followed only by changes of its $m$, $q$ and $j$ as appropriate, but there is
no unique way to reconstruct from these changes how much baryon number
was lost into the black hole.  The black hole forgets how many baryons it has
swallowed.  In fact, it forgets everything but the energy, charge and angular
momentum it ever acquired.  Wheeler refers to this poverty of characteristics
as baldness: `black holes have no hair' (Ruffini and Wheeler 1971). 

After much evidence was garnered in its support in the 1970's and 80's, black
hole baldness was put in question in the 1990's by the discovery of what is
termed `hairy black holes': solutions representing quiescent black holes more
complicated than the KNBH ones, and sometimes with extra parameters
(Heusler 1996).  I have argued (Bekenstein 1997) that since most of these hairy
black holes are unstable, Wheeler's basic idea remains: a black hole
has just a few parameters, whose number is independent of the black hole size.

\bigskip\centerline{\bf  Black Hole Thermodynamics }
\medskip

A black hole can form from collapse of an extremely complex mess of atoms,
ions, radiation.  Yet it transpires that once this object has settled down to
the only stable available black hole configuration - the KNBH one - it
must be specifiable by just three numbers: $m$, $q$ and $j$.  This is a
paradox.  But we are familiar with a similar situation.  A cup of hot tea is
an agglomeration of trillions of molecules of a number of species, all dancing
around violently.  We know, though, that from a macroscopic point of view we
may describe all there is to describe about the equilibrium situation by
giving energy $E$, volume $V$ and the mole fractions of sucrose, and the
various compounds found in tea essence and in lemon.  A few variables describe
the big mess.  Why not try to describe black holes by thermodynamics given
the essential similarity of the situations ?  This was one of the motivations
for black hole thermodynamics (Bekenstein 1973).

A second paradox played an important role.  In 1971 Wheeler stressed to me - I
was then one of his novice doctoral students - that black holes seem to
provide a mechanism for violating the second law of thermodynamics.  Mix the
cup of hot tea with one of cold water and so create entropy.  Then dump the
lukewarm mix into a black hole.  The newly made entropy disappears permanently
from our sight - for we have no interest to follow it into the black hole and
be lost.  The black hole does change, of course.  But how can we figure out
the amount of lost entropy from the changes in  $m$, $q$ and $j$ ?  Not in a
unique way !  So, Wheeler concluded triumphantly, the perfect crime - erasing
an increase of entropy - has been perpetrated.  My incredulity as to whether
such a useful and universal law as the second law could be so easily brushed
aside sent me into a feverish struggle to straighten out the situation.

My basic idea was to ascribe entropy to a black hole, not the entropy of
matter that has gone down the black hole, but some function of the
``observables''  $m$, $q$ and $j$ with the requisite properties.  And the
central desirable property is, of course, that black hole entropy $S_{\rm BH}$
should tend to grow.  Here a nice candidate presented itself. Christodoulou
(1970) and Floyd and Penrose (1971) had given strong evidence that, at
least in classical physics, the area $A$ of a black hole horizon cannot
decrease. Hawking (1971) made this into a theorem.  I thus concluded that
$S_{\rm BH} = f(A)$ with $f$ real, positive and monotonically increasing in
its argument.  The choice $f(A)={\rm const.}\times\surd A$ beckoned because
it implies for $q=0$ and $j=0$ that $S_{\rm BH} \propto m$ in harmony with the
extensive character of entropy in ordinary thermodynamics.  I rejected this
choice, though, because conservation of energy implies that when two
Schwarzschild black holes $a$ and $b$ merge, the mass of the resulting black
hole  is below $m_a+m_b$ given that some losses to gravitational radiation
should occur.  But with $S_{\rm BH} \propto m$ this would say that the total
black hole entropy would {\it decrease} !  The next obvious choice, $S_{\rm
BH} ={\rm const.}\times A$ exhibits no such problem, and so I adopted it. 
Further, on Wheeler's suggestion I took the proportionality constant of the
order $\ell^{-2}$ with $\ell$ the Planck length $(G\hbar/ c^3)^{1/2}\approx
2\cdot 10^{-33}$ cm (theoretical physicists, unlike chemists and engineers,
measure temperature in energy units, in which case Boltzmann's constant $k$
is unity and entropy dimensionless).

With black hole entropy so defined, I could claim to have resolved Wheeler's
paradox.  We knew from Hawking's theorem that the infall of the tea would cause
an increase in the area of the horizon, i.e. and an increment $\delta S_{\rm
BH}$ of the black hole entropy,.  I claimed that in these and similar
situations  $\delta S_{\rm BH}$ would exceed the ordinary entropy lost to the
black hole.  More precisely, I formulated the generalized second law (GSL):
the sum of black hole entropy and ordinary entropy outside black holes never
decreases.  A number of examples showed the law worked  (Bekenstein 1972b,
1973, 1974).

The above historical argument for the form of black hole entropy evidently has
logical gaps in it; for example, it very much guesses the form of $f(A)$. If
we had to define an entropy for a black hole {\it ab initio\/}, what could we
say about it on purely classical grounds ?  To answer this, take the
differential of $A$ in Eq.~(\ref{area}) and rearrange terms to get
\begin{eqnarray}
d(mc^2)&=&\Theta dA +\Phi dq +\Omega dj
\label{1st_law}
\\
\Theta &\equiv& c^4(2GA)^{-1}(R_+-M)
\label{Theta}
\\
\Phi&\equiv&(R_+^2+a^2)^{-1}R_+ q
\\
\Omega&\equiv&(R_+^2+a^2)^{-1}M^{-1}j
\end{eqnarray} 
What is  the significance of $\Phi$ and $\Omega$ ?  If the black hole is
charged, it is surrounded by an electrical potential.  It is found that as one
approaches the horizon, this potential approaches $\Phi$ which is thus aptly
termed the electric potential of the black hole.  Further, if the black hole
rotates,  it is found that it tends to drag infalling objects around it in the
sense of its rotation.  As each such objects falls to the horizon, its angular
frequency approaches $\Omega$, which is thus appropriately identified as the
rotational angular frequency of the black hole.  

Eq.(\ref{1st_law}) reminds us of the first law of thermodynamics for an object
of energy $E$, charged with potential $\tilde\Phi$ and rotating with frequency
$\tilde\Omega$:
\begin{equation}
dE=TdS + \tilde\Phi dq +\tilde\Omega dj
\label{dE}
\end{equation}
Comparing Eqs.(\ref{1st_law}) and (\ref{dE}) and taking into account the
mass-energy equivalence tells us that if we want to regard a KNBH as a
thermodynamic system and Eq.~(\ref{1st_law}) as the  first law of
thermodynamics for it, we must identify the differential of its entropy
$dS_{\rm BH}$ multiplied by its temperature $T_{\rm BH}$ (whatever that may
mean) with $\Theta dA$.  But then, mathematically, we have no choice but to
take $S_{\rm BH}=f(A)$; no more complicated dependence will do (I owe this
point G. Gour).  Further, obviously $T_{\rm BH}=\Theta/f'(A)$.  Since we want
$T_{\rm BH}$ non-negative, we must take $f'(A) >0$.  Thus we have recovered part
of my original claim by a different, stricter route.  Note that it will not
do, as have a number of people, to redefine black hole entropy as depending on
$m$, $q$ and $j$ in a more complicated way than just through $A$.  This clashes
with the first law !

What can we say about $f$ on classical grounds ?  I introduced it as a 
property of a KNBH, but it actually is more widely
applicable.  Consider slowly lowering towards a black hole with $j=0$ but
$q\neq 0$ from opposite sides two equal massive and possibly charged objects,
or instead lowering a massive symmetric ring lying in the hole's
equatorial plane.  By symmetry the hole stays immobile, but it should
become distorted.  Because the process is slow (adiabatic) it is subject
to a rule that the horizon area will not change (Bekenstein 1998; Mayo
1998).  And the process looks reversible, so the black hole entropy should
not change.  All this indicates that also for a {\it distorted\/} KNBH, $S_{\rm
BH}=f(A)$ with the old $f$.  In other words, stationary distortion does not
introduce extra variables in the black hole entropy of a nonrotating KNBH.  A
similar conclusion can be established for a $j\neq 0$ distorted KNBH.

Now imagine a number of such black holes in vacuum held at rest at some
distance one from the other.  There is no evident source of entropy besides the
horizons, and we understand entropy of independent systems to be additive, so
we can write $S=\sum_i\ f(A_i)$.  We now make a the assumption that the entropy
contributed by a black hole is not changed by its motion and associated
dynamical changes, so that the formula applies also when the black holes fall
toward each other (by Hawking's result, though, the individual $A_i$ are then
on the increase).  Consider two of the black holes,
$a$ and $b$, which fall together and merge into a single one.  Horizon area
does not jump up suddenly, but grows smoothly (Hawking 1971).  Thus at the
``moment'" of merging, the area of the new black holes is $A_{\rm
new}=A_a+A_b$.  Our assumption allows us to use $S_{\rm BH}=f(A)$ for the new
black hole.  In the process of merging, as during the  infall, gravitational
radiation (and perhaps electromagnetic one if the holes are charged) is
emitted.  But this is coherent radiation, and should have negligible entropy. 
Thus we are led to assume that just at merger $f(A_a+A_b)=f(A_a)+f(A_b)$.  We
can verify by differentiating with respect to $A_a$, and taking account that
$A_b$ could be anything, that this necessarily requires $f(A)\propto A$ with
no possibility of adding a constant (as one can do, classically, for
entropy).  Thus the  area law of black hole entropy is required by classical
physics.  

But what about the coefficient of the area in this law ?  This seems to require
quantum physics for its determination.  I mentioned Wheeler's suggestion that
the right order of magnitude of $S_{\rm BH}$ should be gotten if we just divide
$A$ by the Planck length squared $\ell^2$.  Support for this came from the
observation (Bekenstein 1973) that when an elementary particle (a thing whose
dimension is of order its Compton length) is very softly deposited at the
horizon of any KNBH, the minimal increase of $S_{\rm BH}$ as just calibrated,
is of order unity.  Since an elementary particle should carry no more than a
unit or so of entropy, the GSL would work, but it would not if we took the
$\ell$ to be much larger than Planck's length. Also it made no sense to take
$\ell$ smaller than Planck's length, which is regarded as the smallest scale on
which smooth spacetime (a must for the black hole concept) is a reasonable
paradigm.   Nothing could be said about the pure numerical coefficient
$\eta$ in the proposed formula $S_{\rm BH}=\eta \ell^{-2} A$, which,
incidentally, established the black hole temperature I mentioned before as
$T_{\rm BH}=\eta^{-1}\ell^2\Theta$.  What this temperature meant operationally
was not clear, though I did study the matter in detail (Bekenstein 1973,
1974).

Notice the restriction $Q^2+a^2\leq M^2$ required for the expression for
horizon area (\ref{area}) to make sense.  Black holes that just saturate this
limit are termed `extremal'.  Since $\Theta$ in (\ref{Theta}) vanishes for the
extremal black holes, all these have $T_{\rm BH}=0$.  However, $S_{\rm BH}$
does not vanish: the Nernst-Simon statement of the third law of thermodynamics
is thus violated by black holes. However, all evidence is consistent with the
conclusion that the unattainability statement ($T=0$ cannot be reached by a
finite chain of operations) is obeyed.  For black holes the two statements
are not equivalent !
 
\bigskip\centerline{\bf  Hawking Radiation}
\medskip

Hawking had been a leader of the vociferous opposition to black hole
thermodynamics.  In a joint paper, ``The Four Laws of Black Hole
Mechanics'', Bardeen, Carter and Hawking (1973) argued {\it against\/} a
thermodynamic interpretation of formulae like (\ref{1st_law}) and in favor of
a purely mechanical one.  Ironically, many uninformed authors still cite that
paper as one of the sources of black hole thermodynamics !  By his own account
Hawking (1988) was trying to discredit black hole thermodynamics when he set
out to investigate the behavior of quantum fields in the gravitational field
of a spherical body which is collapsing to a Schwarzschild black hole.  To his
surprise he found (Hawking 1974) that the incipient black hole is in a
radiating state, emitting spontaneously and steadily all types of radiation
in nature with a thermal spectrum (basically Planck or Fermi according to
whether bosons or fermions are emitted).  A black hole is hot !  

How can a black hole possibly radiate ?  Is it not defined as a region out of
which nothing can come out ?  Hawking also provided an intuitive explanation
of why no such problem exists.  In the  strong gravitation vicinity of the 
incipient horizon, pairs of quanta are created from vacuum, just as a
electron-positron pairs are created from vacuum in a sufficiently strong
electric field.  And because of the  strong gravitation, one member of a pair
may have negative energy overall (counting its big negative gravitational
binding energy).  Since negative energy particles cannot exist in the familiar
spacetime far from a black hole, those pair members have an ephemeral lifetime
and soon enough get swallowed by the hole.  Their partners, though, necessarily
have positive energies (since each pair came out of nothing with zero total
energy); they may thus escape to large distance, and those that do constitute
the Hawking radiation.  It is pretty clear that Hawking emission is a quantum
process.   

Hawking's result, rederived over the years by a number of workers using a
score of different approaches, left him no choice but to accept the  verity of
black hole thermodynamics. The temperature of Hawking's radiance, the same
for all species of quanta, came out to be $4\ell^2\Theta$.  This jibed with
my proposal for black hole temperature provided $\eta={\scriptstyle 1\over
\scriptstyle 4}$.    A large number of calculations since have verified this
value for the black hole temperature.  Hawking's work thus served to
calibrate the black hole entropy formula.

Hawking's justly acclaimed discovery also proffered a striking confirmation of
the GSL.  As used up to that point, this law was checked
only for situations when ordinary entropy is lost into a black hole, being
overcompensated by a growth in black hole entropy.  But Hawking's radiance
disclosed an unexpected possibility.  The Hawking radiation slowly drain
the black hole's mass energy $m$.  For a Schwarzschild hole, $S_{\rm
BH}\propto m^2$, so the black hole entropy {\it decreases}.  The only way the
GSL can hold, then, is for the emergent radiation to carry enough entropy to
overcompensate the loss.  And it does, as checked by Hawking (1976) and by
me (Bekenstein 1975).  In 1971 when the GSL was formulated, the Hawking
radiance was undreamed of.  The just described success thus amounts to
confirmation of a {\it bona fide\/} prediction.  Successes such as this have
made the GSL a pillar of black hole physics in the eyes of gravity
and string theorists. 

\bigskip\centerline{\bf  Black Hole Statistical Mechanics ?}
\medskip

In statistical mechanics, the entropy of an ordinary object is
a measure of the number of states available to it, for example, the logarithm
of the number of quantum states that it may access given its energy.   This
is the statistical meaning of entropy.  What, in this sense, does black hole
entropy represent ?  Is there a black hole statistical mechanics ?

Black hole entropy is large; for instance, a solar mass black hole has $S_{\rm
BH}\approx 10^{79}$ whereas the sun has $S\approx 10^{57}$.  Early on I
expressed the view (Bekenstein 1973, 1975) that black hole entropy is the
logarithm of the number of quantum configurations of {\it any matter\/} that
could have served as its origin.  The `any matter' qualification is
in entire harmony with the `no hair' principle.  A Schwarzschild black hole
of mass $m$ can have come from a mass $m$ of atomic hydrogen, or a mass $m$
of electron-positron plasma, or a mass $m$ of photons, or for that matter any
combination of these and other compositions adding up to mass $m$.  One
cannot distinguish the various possibilities by measuring anything about the
hole.  There is obviously much more entropy here (many more possible states)
than in a mass $m$ of a pure `substance'.  This interpretation would go a
long way towards explaining why black hole entropy is large.  But, you
counter, $S_{\rm BH}$ varies like mass squared, not like mass, which is the
reason for its bigness.  So let us compare a black hole of very small mass
with the same mass of ordinary stuff.  Indeed, a black hole of mass
$10^{15}\,$g has an entropy similar to that of $10^{15}\,$g of ordinary
matter.  However, such a black hole is only about $10^{-13}\,$cm across. 
Matter cannot collapse to make it because the hole is no bigger than the
Compton wavelengths of elementary particles.  It transpires that nature would
not allow a black hole to form whose entropy is not large by ``matter
standards".  The suggested interpretation of black hole entropy thus has an
air of self-consistency about it.

But it cannot be the whole truth.  Suppose we start with a Schwarzschild
black hole.  It is, of course, emitting Hawking radiation at some rate, and
would thus lose mass (and entropy) in the course of time.  But suppose
(Fiola, et al 1994) we arrange for a stream of matter to pour into it at just
such a rate as to balance the mass loss, but without adding charge or angular
momentum to the black hole.  So the black hole does not change in time, and
neither does its entropy.  But surely the inflowing matter is bringing into
the black hole fresh quantum states; yet this is not reflected in a growth of
$S_{\rm BH}$ !  True, radiation is streaming out; could it be carrying away
those additional quantum states, or their equivalent ?  If we continue
thinking of the Hawking radiation as originating outside the horizon, this
does not sound possible.  We are left with the realization that the proposed
black hole entropy interpretation is not handling this example well.

An alternative interpretation is that black hole entropy is the entropy of
quantum fluctuations of material fields in the vicinity of the horizon. 
Thorne and Zurek (1985) first proposed this ``quantum atmosphere'' picture,
and the idea was further developed by 't Hooft (1985) and many others.  It
has the advantage that the linear dependence on $S_{\rm BH}$ on horizon area
is automatic.  However, the coefficient $\eta$ in the formula comes out
formally infinite unless one admits that the fluctuations are suppressed in
a layer next to the horizon.  The right order of magnitude of $\eta$ is
obtained if that layer's thickness  is of order of the Planck length. 
Although this scale might have been expected, one can hardly derive the
value of $\eta$ this way.  An added problem is that the entropy would come
out proportional to the number of material fields in nature; different
conceptions of nature would lead to different coefficients, yet Hawking's
original inference of the coefficient ${\scriptstyle 1\over\scriptstyle 4}$
leaves no room for such freedom. 

An improved approach along this line is that of Carlip (1999) and Solodukhin
(1999).  Here the focus is on fluctuations of the gravitational field, not
matter.  The propinquity of the horizon makes these obey the laws of a
conformal field theory in two spatial dimensions (that being the
dimensionality of the horizon).  Now this sort of theory has been thoroughly
investigated, and using that formalism Carlip and Solodukhin show that the
number of states associated with the fluctuations, when translated into an
entropy, is exactly the accepted formula, coefficient and all.

How can we understand intuitively a part of this important result ?  Suppose,
as has been suggested (Bekenstein and Mukhanov 1995), and verified by many
workers in a variety of ways, that the area of the horizon is quantized with
uniformly spaced levels of order of the squared Planck length:
$A=\alpha\ell^2 n$  with $\alpha$ a positive pure number and $n=1,2,\,
\cdots\,$. This suggests that the horizon is to be thought of as a patchwork of patches with area
$\alpha\ell^2$.    If every patch can have, say, 2 distinct states, then
a black hole with area $A=\alpha\ell^2 n$ can be in any of $2^n$ `surface'
states.  As usual, degeneracy makes a contribution to the entropy. 
If there is no other contribution, then $S_{\rm BH}=\ln 2^n=\ln
2\cdot(\alpha\ell^{2})^{-1}A$ and we have recovered the area law (Bekenstein
1999, Sorkin 1998).  Further, from Hawking we know that $\alpha^{-1}\ln 2=
{\scriptstyle 1\over\scriptstyle 4}$ so that the horizon quantization law is
$A=4\ln 2\cdot\ell^2 n$.

This uniform spacing of the horizon area spectrum is not universally
accepted.  The Ashtekar (loop gravity) school of quantum gravity derives a
rather more complicated spectrum.  Surprisingly enough, a statistical mechanics
of the horizon based on this spectrum does recover the law $S_{\rm BH}\propto
A$ law, although without giving the coefficient (Ashtekar, et al. 1998).   Of
late a claim that loop quantum gravity actually does lead to a uniformly
spaced area spectrum has appeared (Alekseev, et al. 2000).  In the face of
such basic disagreement, it is unclear what to make of the just mentioned
black hole statistical mechanics. 

String theorists have in the last years claimed to have completely clarified
the statistical mechanical origin of the formula $S_{\rm BH}={\scriptstyle
1\over\scriptstyle 4}A/\ell^2$.  String theory regards elementary particles
like the electron or the photon as vibrations of more fundamental entities,
the strings.  These are one-dimensional objects which move in higher (10 or
26) dimensional spacetime.  String theory admits other solutions to its
equations, the Dirichlet branes - or branes for short.  Strings are to branes
as cords are to membranes in our humble three dimensional space.  String
theorists have found that arrays of branes can have black hole properties. 
Actually these brane black holes carry parameters different from those of the
Kerr-Newman black hole.  Nevertheless, string theorists are able to identify
what might be called the horizon area $A$ and to establish its dependence on
these new parameters.  And they are able to count the number of different
brane configurations that correspond to a particular $A$.  The logarithm of
this ``degeneracy'' is taken to be the entropy associated with that
set of configurations.  Apart from well understood higher corrections, this
turns out to coincide with ${\scriptstyle 1\over\scriptstyle 4}A/\ell^2$
(Strominger and Vafa 1996).

Despite this triumph, this brand of black hole statistical mechanics
has drawbacks.  First, the program is cleanly executable only for extremal
black holes (some claims to have escaped this restriction have not attained
general recognition), but we already know that extremal black holes are
pathological.  For example, one cannot reach such a black hole state starting
from the other black holes.  Second, it is unclear in what sense the
conglomerations of branes being considered are the same as the black holes
one would find, say, in the aftermath of gravitational collapse in nature. 
When impressed by the reproduction of the black hole entropy formula from
branes, I am thus reminded of the story told about George Gamow, the colorful
Russian-American physicist who with Ralph Alpher predicted in 1948 that the
universe should be full of thermal radiation with temperature 5-10$^0$ K.  In
1965 Penzias and Wilson discovered the celebrated 3$^0$ K microwave background
radiation (for which discovery they later shared the Nobel prize with Ryle). 
Gamow was asked how he felt about this confirmation of his prediction.
``Well'', he retorted, ``if you have lost a nickel, and somebody has found a
nickel, it does not prove it is the same nickel''.  Are brane black holes
and traditional black holes the same nickel ?    

\bigskip\centerline{\bf  The Bounds on Information}
\medskip

How much information can be stored by {\it whatever\/} means in a cube  of
{\it whatever\/} composition one centimeter on the side ?  Foreseeable
technology making use of atomic manipulation would suggest an upper bound of
around $10^{20}$ bits.  But as technology takes advantage of unforeseen
paradigms, this number could - and will - go up.  For example, we might one
day harness the atomic nucleus as an information cache.  Can the bound go up
without limit ?  Thirty years ago we would not have known what to answer. But
with black hole thermodynamics some definite answers are forthcoming.

First, by information theory, the maximal information ${\cal I}_{\rm max}$ a
system  can hold, reachable if we know in detail its state, is
numerically  (up to a factor $\ln 2$) just the maximal entropy $S_{\rm max}$
it could hold under the complementary circumstance that we know nothing about
its internal state.  Now suppose our information cache is coaxed into
collapsing into a black hole.  Obviously its surface area ${\cal A}$ will
shrink in the process.  The resulting black hole has an entropy ${\scriptstyle
1\over\scriptstyle 4}A_{\rm BH}/\ell^2$ with
$A_{\rm BH}<{\cal A}$.  But by the GSL this entropy must exceed
$S_{\rm max}$.  Thus ${\cal I}_{\rm max}\ln 2= S_{\rm max}<S_{\rm
BH}<{\scriptstyle 1\over\scriptstyle 4}{\cal A}/\ell^2$, or
\begin{equation}
{\cal I}_{\rm max}<{{\cal A}\over 4\ell^2\ln 2 }.
\label{holographic}
\end{equation}
This bound was inferred by 't Hooft (1993) and Susskind (1995) by following
very much the previous reasoning; 't Hooft termed it `holographic'.  By now
Eq.(\ref{holographic}) has come to be part and parcel of a whole philosophy -
the holographic principle espoused by string and gravity theorists - of what
constitutes an acceptable physical theory.  But as an information bound, the
holographic one seems exaggerated.  Our standard one centimeter cube is only
required by it to hold no more than some $10^{65}$ bits.  It is hard to
conceive how any technology can ever span the gulf of 45 orders of magnitude
between foreseeable information capacity and this figure.

A more efficient information bound came up already in 1980 from my
turning around the logic that supplied support for the GSL (Bekenstein 1981). 
Suppose we have faith in the validity of the GSL.  Drop an information
cache of overall radius $R$ and total mass-energy $E$ gently into a black hole
so that it causes a minimum of horizon area increase.  This minimum
is determined by purely mechanical considerations; it is $8\pi GREc^{-4}$. 
In accordance with the GSL, demand now that the corresponding increase in
black hole entropy be no smaller than the maximum entropy the cache can hold,
namely ${\cal I}_{\rm max}\ln 2$.  The result is the bound (Bekenstein 1981)
\begin{equation}
{\cal I}_{\rm max}<{2\pi RE\over c\hbar\ln 2 }
\label{bound}
\end{equation}
Unlike (\ref{holographic}), this bound does not contain $G$; it looks very
`everyday' indeed.  In fact, for simple closed systems (\ref{bound}) can
also be derived from quantum statistical considerations without even
mentioning black holes (Bekenstein and Schiffer 1990).  And (\ref{bound})
is generally a tighter bound than (\ref{holographic}).  For instance, it
requires our one centimeter cube, if made of ordinary materials, to hold no
more than about $10^{38}$ bits.  This is 27 orders of magnitude tighter than
the holographic requirement, and ``only'' 18 orders above the foreseeable
information capacity.   I believe better bounds can be found.

The logic leading to both of the above bounds obviously assumes that the
ordinary entropy mentioned in the GSL is the {\it total\/} entropy of the
system at all levels.  As we know, were we to ignore the atomic, nuclear,
quark, and possibly deeper degrees of freedom of matter, and compute the
entropy of its molecules by statistical mechanics, we would miss out
contributions to the total entropy.  How do we know that the GSL ``sees'' all
these ?  Because it is a gravitational law (for example, $G$ appears in the
black hole entropy contribution), and gravitation, unlike other interactions,
is aware of all degrees of freedom because they all gravitate.  In the end
this is what allows black hole thermodynamics to inform us about subtle
aspects of ordinary physical systems, like the limits of information.      
    
\vspace{0.5cm}

{\bf Acknowledgment:} I thank Avraham Mayo and Gilad Gour for many
conversations; this research is supported by the Hebrew
University's Intramural Fund. 
\vspace{0.5cm}
\parskip=5pt

\centerline{\bf Bibliography}
\vspace{0.3cm}

\noindent
Alekseev, A.,  Polychronakos, A. P. and Smedb\"ack, M. (2000) On area and
entropy of a black hole, Los Alamos Archives preprint hep-th/0004036.

\noindent
Ashtekar, A.,  Baez, J.,  Corichi, A. and Krasnov, K. (1998) Quantum
Geometry and Black Hole Entropy, Physical Review Letters 80, 904-907.

\noindent
Bardeen, J. M., Carter, B. and Hawking, S. W. (1973), The Four Laws of Black
Hole Mechanics, Communications of Mathematical Physics 31, 161-170. 

\noindent
Bekenstein, J. D. (1972a) Transcendence of the Law of Baryon Number
Conservation in Black Hole Physics, Physical Review Letters  28, 452-455.

\noindent
Bekenstein, J. D. (1972b) Black Holes and the Second Law, Lettere al 
Nuovo Cimento 4, 737-740.

\noindent
Bekenstein, J. D. (1973) Black Holes and Entropy, Physical Review
D 7, 2333-2346.

\noindent
Bekenstein, J. D. (1974) Generalized Second Law of Thermodynamics in Black
Hole Physics, Physical Review D 9, 3292-3300.

\noindent
Bekenstein, J. D. (1975) Statistical Black Hole Thermodynamics, 
Physical Review D 12, 3077-3085.

\noindent
Bekenstein, J. D. (1981) Universal Upper  Bound on Entropy to Energy Ratio
for Bounded Systems,  Physical Review D 23, 287-298.

\noindent
Bekenstein, J. D. and Schiffer, M. S. (1990) Quantum Limitations on the
Storage and Transmission of Information, International Journal of Modern
Physics C 1, 355-422.

\noindent
Bekenstein J. D. and Mukhanov, V. F. (1995) Spectroscopy of the Quantum
Black Hole, Physics Letters B 360, 7-12. 

\noindent
Bekenstein, J. D. (1997) `No Hair': Twenty--five Years
After, in I. M. Dremin and A. Semikhatov (eds),  Proceedings of the  Second
Andrei D. Sakharov Conference in Physics (Singapore: World Scientific).

\noindent
Bekenstein, J. D. (1998) Disturbing the Black Hole, in  B. Iyer and
B. Bhawal (eds)  Black Holes, Gravitational Radiation and the
Universe (Dordrecht: Kluwer) pp. 87-101.

\noindent
Bekenstein, J. D. (1999) Quantum Black Holes as Atoms, in T. Piran and
R. Ruffini (eds), Proceedings of the Eight Marcel Grossman Meeting on General
Relativity,  (Singapore: World Scientific) pp. 92-111. 

\noindent
Carlip, S. (1999) Black Hole Entropy from Conformal Field Theory in Any
Dimension, Physical Review Letters 82, 2828-2831.

\noindent
Christodoulou D. (1970) Reversible and Irreversible Transformations
in Black Hole Physics, Physical Review Letters  25, 1596-1597. 

\noindent
Fiola, T. M., Preskill, J., Strominger, A. and Trivedi, S. P. (1994) Black
Hole Thermodynamics and Information Loss in Two Dimensions, Physical Review
D 50, 3987-4014.

\noindent
Floyd, R. M. and Penrose, R. (1971) Extraction of Rotational Energy from
a Black Hole, Nature (Physical Science) 229, 177-179.

\noindent
Hawking S. W. (1971) Gravitational Radiation from Colliding Black Holes, 
Physical Review Letters  26, 1344-1346.

\noindent
Hawking, S. W. (1974) Particle Creation by Black Holes, Communications of
Mathematical Physics 43, 199-220.

\noindent
Hawking S. W. (1976)  Black Holes and Thermodynamics, Physical Review D 13, 
191-197. 

\noindent
Hawking, S. W. (1988)  A Brief History of Time (Toronto: Bantam Books
1988).

\noindent
Heusler, M. (1996)  Black Hole Uniqueness Theorems (Cambridge:
Cambridge University Press).

\noindent
Misner, C. W., Thorne, K. S. and Wheeler, J. A. (1973)  Gravitation
(San Francisco: Freeman).

\noindent
Mayo, A. E. (1998) Evidence for the Adiabatic Invariance of the Black Hole
Horizon Area, Physical Review D 58, 104007.

\noindent
Ruffini R. and Wheeler J. A. (1971) Introducing the Black Hole, Physics
Today 24, 30-41.

\noindent
Solodukhin, S. (1999) Conformal description of horizon's states, Physics
Letters B 454, 213-222.

\noindent
Sorkin, R. D.(1998) The Statistical Mechanics of Black Hole Thermodynamics, in
Wald, R. M. (ed), Black Holes and Relativistic Stars (Chicago: University of
Chicago Press).

\noindent
Strominger, A. and Vafa, C. (1996) Microscopic Origin of the
Bekenstein-Hawking Entropy, Physics Letters B 379, 99-104.  

\noindent
Susskind, L. (1995) The World as a Hologram, Journal of Mathematical Physics.
36, 6377-6396.

\noindent
't Hooft, G. (1985) On the Quantum Structure of a Black Hole, Nuclear
Physics B 256, 727-745.

\noindent
't Hooft, G. (1993) Dimensional Reduction in Quantum Gravity, in  A. Aly,
J. Ellis and S. Randjbar--Daemi (eds),  Salam--festschrifft (Singapore:
World Scientific).

\noindent
Thorne, K. S. and Zurek, W. H. (1985) Statistical Mechanical Origin of the
Entropy of a Rotating Charged Black Hole,  Physical Review Letters 54,
2171-2175.
 
\end{document}